\documentclass[aps,showpacs,amsmath,amssymb]{revtex4}
\usepackage{graphicx}
\begin{document}
\title{New physical principles of contact thermoelectric cooling}
\author{Yu.~G.~Gurevich}
\email{gurevich@fis.cinvestav.mx}
\affiliation{Depto.~de~F\'{\i}sica, CINVESTAV---IPN,\\
Apdo.~Postal 14--740, 07000 M\'{e}xico, D.F., M\'{e}xico}
\author{G.~N.~Logvinov}
\email{logvinov@fis.cinvestav.mx}
\affiliation{Escuela~de~F\'{\i}sica, Instituto Polit\'{e}cnico 
Nacional, M\'{e}xico, D.F., M\'{e}xico}
\author{O.~Yu.~Titov}
\email{oleg.titov@aleph-tec.com}
\affiliation{CICATA---IPN, Av.~Jos\'{e} Siurob 10, Col.~Alameda,\\
76040 Quer\'{e}taro, Qro., M\'{e}xico}
\author{J.~Giraldo}
\email{jgiraldo@ciencias.unal.edu.co}
\affiliation{Grupo de F\'{\i}sica de la Materia Condensada,\\
Universidad Nacional de Colombia, A.A.60739, Bogot\'{a}, Colombia}
\begin{abstract}
We suggest a new approach to the theory of the contact thermoelectric 
cooling (Peltier effect). The metal-metal, metal-n-type semiconductor, 
metal-p-type semiconductor, p-n junction contacts are analyzed. Both 
degenerate and non-degenerate electron and hole gases are considered. The 
role of recombination in the contact cooling effect is discussed by the 
first time.
\end{abstract}



\maketitle

\section{Introduction}

It is well known \cite{1} that the Peltier effect is the basis for solid
thermoelectric cooling. The main point of it arises from the fact that the
heat extraction or absorption occurs at the contact between two different
conducting media if the d.c. electric current flows through this contact.
Boundary conditions in an electric current contact have been discussed in
a previous work. \cite{2} Here we are interested in the heat flow. The
absorption or extraction of heat (cooling or heating) depends on the
current direction. The traditional approach to determine the quantity of
this heat (Peltier heat) is based on the generalized thermal conductivity
law, \cite{3} \begin{equation} {\mathbf {q}} = -\kappa\nabla T +
(\tilde\mu + \Pi) \mathbf{j} \end{equation} where $\mathbf q$ is the heat
flux density, $\kappa$ is the thermal conductivity, $T$ is the
temperature, $\tilde\mu = \varphi + {\mu}/{e_0}$ is the electrochemical
potential of the charge carriers, $\varphi$ is the electric potential,
$\mu$ is the chemical potential of the carriers, $e_0$ is the charge of
the carriers, $\Pi$ is the Peltier coefficient, $\mathbf j$ is the
electric current density.

Let us suppose that the temperature gradient is absent in the electric
circuit, and that the electric current flows through the contact of two
media (Fig.~1). We assume also that the lateral sides of the circuit are
adiabatically insulated, and that the circuit has a unit section.
\begin{figure}[!ht]
\centering
\includegraphics[width=7cm]{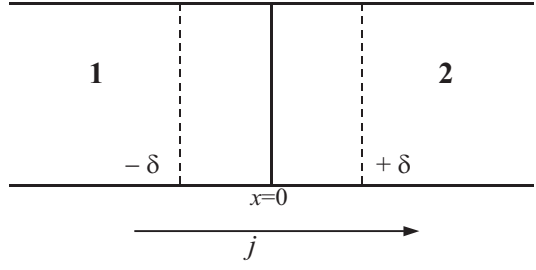}
\caption{Junction for the Peltier effect.}
\label{fig:1}
\end{figure}

Let us choose the small volume bounded by the planes $x = \pm\delta$ near
the contact $x = 0$, and calculate the Peltier heat $Q_{\Pi}$ at this
volume under the presence of the electric current (see Fig.1). In the
linear approximation to the electric current (we neglect the Joule
heating) this heat per unit time is equal to \begin{equation} Q_{\Pi}=
-\int_V\nabla\centerdot {\mathbf {q}}_{\Pi}dv = -\oint\limits_{S} {\mathbf
{q}}_{\Pi}\centerdot d{\mathbf{s}}. \end{equation}

We integrate over the surface limiting the chosen volume in Fig.1.
${\mathbf {q}}_{\Pi}=\Pi {\mathbf {j}}$ is the Peltier heat density,
$d{\mathbf {s}}=ds{\mathbf {n}}$, $ds$ is the surface element, $\mathbf n$
is the external normal. For the simple chosen geometry it is
straightforward to verify that in the limit $\delta \to 0$,

\begin{equation}
Q_{\Pi}=(\Pi_1 - \Pi_2)J,
\end{equation}
where $\Pi_{1,2}$ are the Peltier coefficients of the first and the 
second materials, and  $J$ is the electric current. The quantity  $ 
Q_{\Pi}$ determines the Peltier heat at the contact between the two 
media. Under fixed current direction $Q_{\Pi} > 0$ if  $\Pi_1 > \Pi_2$  
and the contact is heated. On the contrary, if 
$\Pi_1 < \Pi_2$, $Q_{\Pi} < 0$ and the contact is cooled.
Actually the thermoelectric cooling modules are fabricated from two 
semiconductor brunches with electron (n) and hole (p) type 
conductivities.\cite{1} Figure~2 illustrates the sketch of this module. 
Two semiconductors are coupled electrically in series and thermally in 
parallel.
For  non-degenerate electrons (n) and holes (p),\cite{4}
\begin{equation}
\Pi_{n,p} = \mp\frac{1}{e}\left [(q_{n,p} + \frac{5}{2})T-\mu_{n,p}\right].
\end{equation}
Here $T$ is the temperature in energy units, $\mu_{n,p}$ are the chemical
potentials of electrons and holes which are measured from the bottom of
the conduction band and the top of the valence band, respectively ($\mu_p
= -\epsilon_g - \mu_n$, where $\epsilon_g$ is the band gap), $e$ is the
hole charge, $q_{n,p}$ are the exponents in the momentum relaxation
times,\cite{5}

\begin{equation}
\tau_{n,p}(\epsilon) = \tau_{n,p}^ {0}\left (\frac{\epsilon}{T}\right)^{q_{n,p}},
\end{equation}
where $\epsilon$ is the energy of the carriers. The constant quantities $\tau_{n,p}^ {0}$ and $q_{n,p}$ for different relaxation mechanisms can be found elsewhere.\cite{5}.

\begin{figure}[!ht]
\centering
\includegraphics[width=5cm]{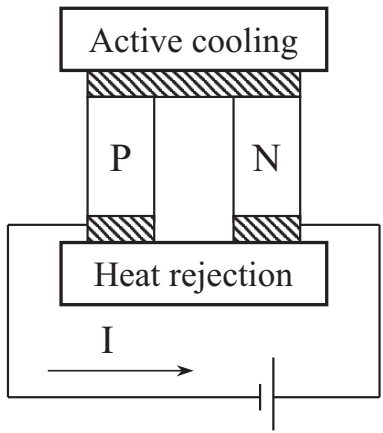}
\caption{Sketch of a thermoelectric cooling module.}
\label{fig:2}
\end{figure}

For degenerate electron and hole gases, 
\begin{equation}
\Pi_{n,p} = \mp\frac{\pi^2}{3e}\left(q_{n,p} + \frac{3}{2}\right)\frac{T^2}{\mu_{n,p}}.
\end{equation}

To understand the physical meaning of equations (4) and (6), it is convenient to rewrite these expressions as follows:

\begin{equation}
\Pi_{n,p} = \mp\frac{1}{e}\left[\langle \epsilon \rangle_{n,p}  -\mu_{n,p}\right],
\end{equation}
where $\langle \epsilon \rangle_{n,p}$ is the mean kinetic energy of the carriers in the flux,

\begin{equation}
\langle \epsilon\rangle_{n,p} = \frac{\int_0^{\infty} \tau_{n,p}(\epsilon) \epsilon^{5/2} 
\frac{df_0^{n,p} (\epsilon)}{d\epsilon} d\epsilon}
{\int_0^{\infty} \tau_{n,p}(\epsilon) \epsilon^{3/2} 
\frac{df_0^{n,p} (\epsilon)}{d\epsilon} d\epsilon}.
\end{equation}
where $f_0^{n,p}(\epsilon)$ are the Fermi-Dirac distribution functions. Notice that, similar to the chemical potentials, the mean energies 
$\langle \epsilon \rangle_{n,p}$ are measured from the bottom of the conduction band and the top of the valence band respectively. Often the chemical potentials $\mu_{n,p}$ are interpreted as the potential energies of electrons and holes.\cite{4}

For non-degenerate statistics ($\vert \mu \vert \gg T, \mu < 0$),

\begin{equation}
\langle \epsilon\rangle = \left (q + \frac{5}{2}\right) T,
\end{equation}
while for the degenerate case ($\vert \mu \vert \gg T, \mu > 0$),
\begin{equation}
\langle \epsilon\rangle = \mu + \frac{\pi^2T^2}{3\mu} \left(q + \frac{3}{2}\right).
\end{equation}

To demonstrate the physical meaning of the Peltier effect,  a metal-n-type semiconductor contact is usually used (see Fig.3). It is easy to see that the surplus of the electron energy in the semiconductor, in comparison with the metal, is	$\Delta \epsilon = \langle \epsilon\rangle - \mu_n  = (q + 5/2) T - \mu_n$, and $\Pi_1 - \Pi_2 = \Delta \epsilon /e$.

\begin{figure}[!ht]
\centering
\includegraphics{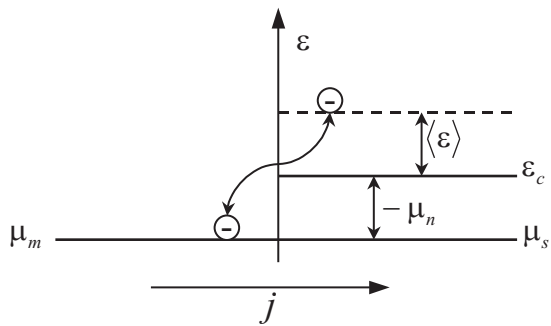}
\caption{Heating of metal-n-type semiconductor contact.}
\label{fig:3}
\end{figure}

The obtained result is clear for this type of contact. At the same time these contacts are not really used in applications. The consideration of other contacts do not have the same obvious understanding, and a priori the following questions arise:
\begin{itemize}
\item What does the potential energy for the degenerate gas of the charge carriers mean?
\item What is the physical meaning of the energy change in a metal-p-semiconductor contact and in a p-n junction?
\item Why does the approach stated above ignore the work of the built-in electric field within the contact?
\item It is easy to show that the value $\langle \epsilon\rangle - \mu$ in any type of contacts is different on the left and on the right from the contact plane. Therefore, the description of the contact refrigeration based on the consideration of the separate electron transitions is not valid. One needs to take into account the statistics of these transitions. How to reflect this fact in the preceding scheme? 
\end{itemize}

\vspace*{-4pt}
\section{Mechanisms of contact thermoelectric cooling}
\noindent
In the usual scheme of n-n or p-p junctions the contact heating or cooling is explained by the electron (hole) transitions between $\langle \epsilon \rangle$ - levels (see Fig.4). The latter are counted from the common Fermi level. 
In p-n junctions it is necessary to take into account the electron-hole generation and recombination. These are the processes that determine contact cooling and heating. In most works devoted to discuss this problem, the recombination rate is assumed to be infinite.

\begin{figure}[!ht]
\centering
\includegraphics{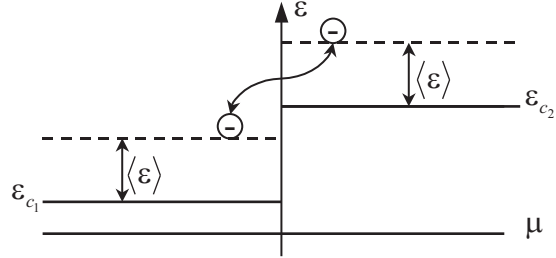}
\caption{Contact in equilibrium ($j = 0$)}.
\label{fig:4}
\end{figure}

Here we present some new assumptions. Let us consider first the n-n contact between  non-degenerate semiconductors (Fig.5). If the momentum scattering in both semiconductors is the same, then [see Eq.(7)],

\begin{eqnarray}
\Pi_1 - \Pi_2 &=& \frac{(\mu_1 - \mu_2)}{e}\\ &=& \frac{(\chi_1^0 - \chi_1 - \chi_2^0 + \chi_2)}{e}
 = \varphi_c + \frac{\Delta \epsilon_c}{e},\nonumber
\end{eqnarray}
where $\varphi_c = (\chi_2 -\chi_1)/e$ is the contact voltage, and $\Delta \epsilon_c = \chi_1^0 -\chi_2^0$ is the work of a valence force. This means that the contact cooling or heating is connected with the work of the built-in electric field $W_c$ and the work of the valence forces $W_{\nu}$. The total work is $W = Q_ {\Pi}= W_c + W_{\nu}$,  where  $W_c = J\varphi_c$  and $W_{\nu} = J \Delta \epsilon_c/e$. Here the valence force is caused by the abrupt change of the conduction band. A similar reasoning is valid for the p-p junction. 

\begin{figure}[!ht]
\centering
\includegraphics[width=8cm]{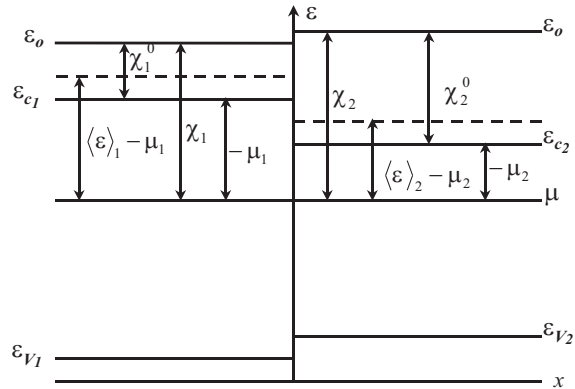}
\caption{Contact of non-degnerate semiconductors. $\epsilon_0$ is the vacuum level; $\chi_i$ and $\chi_{i^0}$ are workfunctions and electron affinities, respectively.}
\label{fig:5}
\end{figure}

If we have the $n^+ -n$ or $p^+ - p$ junction, the Peltier effect does not depend on the contact properties, and it is determined only by the parameters of the non-degenerate semiconductor (see Fig.3),
\begin{equation}
\Pi_1 - \Pi_2 = -\frac{\mu_n}{e}. 
\end{equation}

In the case of a contact between two degenerate n- or p-type semiconductors (or two metals) the Peltier effect depends on the contact properties again (see Figs.6 and Eq.(10)), and the work of the built-in electric field jointly with the work of the valence forces occurs. 
Now
\begin{equation}
\Pi_1 - \Pi_2 = \frac{\pi^2}{3e} \left (q + \frac{3}{2}\right) \frac {T^2}{\mu_1 \mu_2}(\mu_1 - \mu_2). 
\end{equation}

\begin{figure}[!ht]
\centering
\includegraphics[width=8cm]{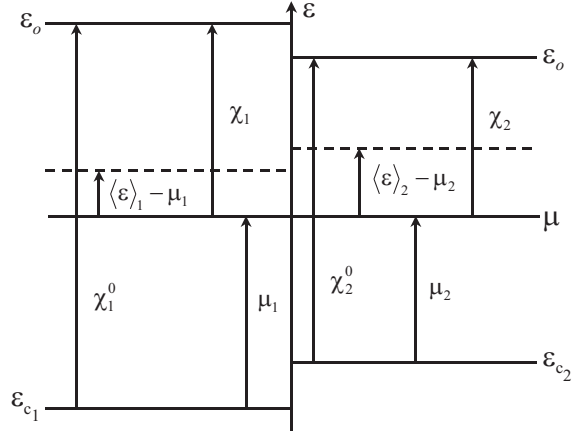}
\caption{A contact between two metals.}
\label{fig:6}
\end{figure}

\begin{figure}[!ht]
\centering
\includegraphics[width=8cm]{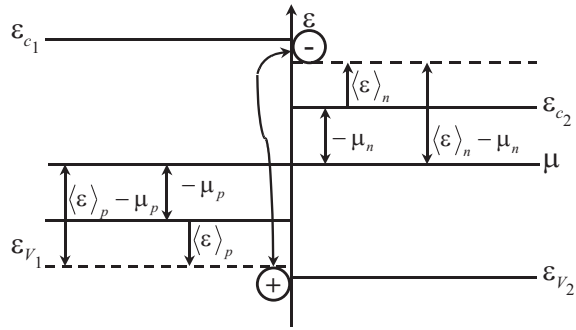}
\caption{A p-n junction.}
\label{fig:7}
\end{figure}

Comparing (10) and (12) we see that the Peltier heat in the contact between degenerate materials is essentially less than that in the contact between the non-degenerate semiconductors. 

Let us note also that the work of the built-in electric field and the valence forces give rise to the qualitatively different temperature of the contacts in the case of non-degenerate and degenerate brunches of the circuit. In the former
\begin{equation}
W_c \propto \frac {3}{2} (T(E) - T_0). 
\end{equation}
where  $ T(E)$ is the non-equilibrium temperature, $E$ is the built-in electric field, $ T_0$ is the room temperature. In the latter,
\begin{equation}
W_c \propto \frac{\pi^2}{4} \frac {T^2(E) - T_0^2}{\mu}. 
\end{equation}

Let us examine the p-n junction (Fig.7). It is easy to see from Fig.7 that
\begin{equation}
\Pi_1 - \Pi_2 = - \frac{\mu_n + \mu_p}{e}. 
\end{equation}

The generation or recombination at the contact causes the Peltier heating or cooling in this case. In addition, Eq.(15) corresponds to the infinitely strong recombination rate. At the same time there are a lot of situations when this rate is weak enough. What has happened with the Peltier effect in this case? The situation is similar to the electric current or thermoelectric current in the bipolar semiconductors. \cite{6,7} The Fermi quasilevels must appear causing the redistribution of carriers concentrations, electric fields and conductivity. As a result the nature of the Peltier effect can drastically change. It follows from Fig.8 that under the chosen current direction the cooling exchanges to the heating if the recombination is absent. We do not present here the detailed calculations, and give only the main equations for this problem.

\begin{figure}[!ht]
\centering
\includegraphics[width=8cm]{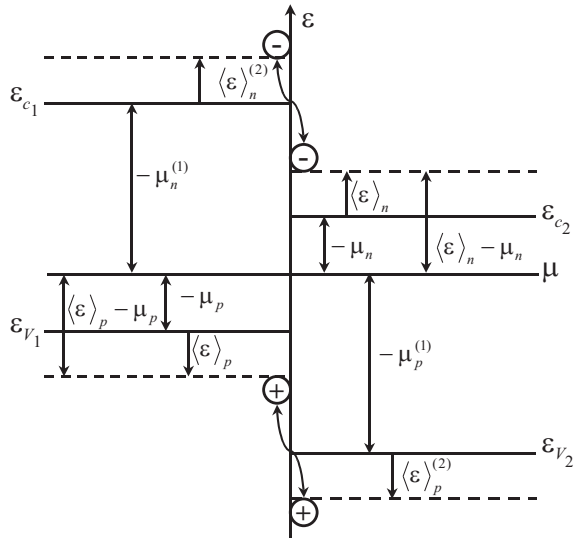}
\caption{ {\it A p-n junction without recombination}.}
\label{fig:8}
\end{figure}

\section{Main equations}
The general approach to the problem of the
contact cooling or heating accounting for the generation and recombination
processes requires the solution of the following set of equations:

\begin{eqnarray}
\nabla \centerdot {\mathbf {j}}_n = eR_n,
\nabla \centerdot {\mathbf {j}}_p = -eR_p,\\
\nabla \varphi(x) = 4\pi e(\delta n - \delta p).\nonumber
\end{eqnarray}

Here $\mathbf {j_{n,p}}$ are the electron and hole electric current densities; $\delta n$, $\delta p$ are the concentrations of non-equilibrium electrons and holes, and $R_{n,p}$ are the electron and hole recombination rates.\\
A rigorous theory of recombination suggests that $R_n$ is identically equal to $R_p$ and therefore for all recombinatiion mechanisms in a linear approximation we have \cite{2,8}
\begin{equation}
R_n = R_p = R = \frac{\delta n}{\tau_n} + \frac{\delta p}{\tau_p}.
\end{equation}

The values of  $\tau_{n,p}$ are material parameters having the dimension of time, but they are not electron and hole lifetimes. However, when the condition of quasi-neutrality holds \cite{9} ($r_d \ll a^2$, where  $r_d$ is the Debye radius and  $a$ is the width of the p-n junction), as it is often the case, the non-equilibrium electron and hole concentrations are equal ($\delta n = \delta p$). In this case one can define a lifetime  common to electrons and holes as

\begin{equation}
\frac{1}{\tau} = \frac{1}{\tau_n} + \frac{1}{\tau_p}.
\end{equation}

From a previous work,\cite{8} it follows that $\tau_n /\tau_p = n_0/p_0$; 
therefore, the lifetime of non-equilibrium carriers can be expressed as 
\begin{equation}
\tau = \tau_n \frac{p_o}{n_0 + p_0}.
\end{equation}

As it is well known, \cite{9} using Poisson equation in a quasi-neutrality approximation becomes unnecessary, and the electric potential at the boundary of two semiconductors displays jump-like changes. The boundary conditions to Eqs.(16)-(18) can be found elsewhere. \cite{2} The surface recombination rates present in these boundary conditions are \cite{8,10}
\begin{equation}
R_s = S_n \delta n + S_p \delta p, 
\end{equation}
where $S_n$ and $S_p$ are the parameters characterizing the contact properties. If the quasi-neutrality conditions ($\delta n = \delta p$) are satisfied, a surface recombination rate common to electrons and holes can be introduced:
\begin{equation}
 S = S_n + S_p.
\end{equation}
Therefore one can write
\begin{equation}
R_s = S \delta n.
\end{equation}

\section{Further discussion and conclusions}
In this section we will present qualitative arguments regarding the two limiting cases of very strong and very weak recombination rates. The criteria for these rates have been discussed in \cite{6}. According to those results, the surface and the bulk recombination rates are strong if $S \gg S_0$, or $\tau_n \ll \tau_0$ respectively, where $S_0 = (n_0 + p_0) \sigma_p^2T_0 / (\sigma_n + \sigma_p) e^2an_0p_0$,  $n_0$ and $p_0$ are the equilibrium concentration of electrons and holes, $\sigma_n$ and $\sigma_p$ and  are the electron and hole conductivity; $\tau_0 = e^2an_0T_0(\sigma_n + \sigma_p) / (\sigma_n  \sigma_p)$.  The inequalities  $S \ll S_0$ and $\tau_n \gg \tau_0$ determine the weak surface and bulk recombination. 
If the strong recombination occurs, then the physical process in the p-n junction is as in Fig.7, and Eq. (12) is right. In the opposite case (weak recombination), electrons from the n-region will pass to the p-region and occupy the states with  mean energy  ${\langle \epsilon \rangle_n}$ (cf. Fig.8). By analogy holes will pass to the n-region and will occupy the states with the mean energy ${\langle \epsilon \rangle_p}$. In the latter case, under the appropriate current direction, the contact is heated and the cooling is absent.

In conclusion, we wish to emphasize that the contact refrigeration is a promising field for applications in different spheres of human activity. In particular, a proper description of the physical processes involved is very important in the today's problem of acheaving high values of the figure of merit  $Z$ for thermoelectric materials. \cite{11} All the efforts devoted to this problem up to now \cite{12} have been based on the expression  $Z = (\alpha_n + \alpha_p)^2 / (\sqrt {\kappa_n \rho_n} + \sqrt {\kappa_p \rho_p})^2$ \cite{1}or in the simpler expression $Z = \alpha^2\sigma / \kappa$. Here $\alpha_{n,p}$, $\kappa_{n,p}$ and $\rho_{n,p}$ are, respectively, electron and hole thermoelectric power, thermal conductivity and electric resistance. $\alpha$ and  $\sigma$ are the thermoelectric power and the electric conductivity of an effective medium.

We want to underline that in thermoelectricity this expression is good for the circuit consisting of the metal and semiconductor of n-type and for another materials if the surface and the bulk recombination is very strong. At the same time this formula must be revised in the case of the contacts metal-p-semiconductor and in p-n junctions, where the recombination processes are essential even in the linear approximation. \cite{6,7} Especially important is to take into account the recombination processes in p-n junctions in thin-film thermoelectric cooling modules (Fig.2).

This work was partially supported by CONACyT, M\'{e}xico. J.G. aknowledges support from DINAIN,  Universidad Nacional de Colombia, Colombia.

\end{document}